\documentclass[twocolumn,showpacs]{revtex4}
\def\address{\affiliation}
\usepackage[dvips]{graphicx}
\begin{document}

\title{
Thermoelectric properties of the brownmillerite oxide
Ca$_{2-y}$La$_y$Co$_{2-x}$Al$_x$O$_5$
}

\author{
Wataru {\sc Kobayashi}$^1$, Akira {\sc Satake}$^1$ and 
Ichiro {\sc Terasaki}$^{1,2}$\footnote{E-mail address:terra@mn.waseda.ac.jp}
}

\address{
$^{1}$Department of Applied Physics, Waseda University,
3--4--1 Ohkubo, Shinjuku-ku, Tokyo, 169-8555, Japan\\
$^{2}$Precursory Research for Embryonic Science and
Technology, Japan Science Technology, Tokyo 108-0075, Japan
}

\begin{abstract}
We prepared the brownmillerite oxide Ca$_{2-y}$La$_y$Co$_{2-x}$Al$_x$O$_5$, 
and found that it was an n-type conductor. 
The thermopower and the resistivity of the single crystal are $-$90~$\mu$V/K 
and 68~m$\Omega$cm along the $ab$ direction at 440~K, 
which suggest relatively good thermoelectrical properties,
compared with other transition-metal oxides. 
Their temperature dependences are of activation type, 
and the activation energies are 0.2~eV for the resistivity 
and 0.04~eV for the thermopower. 
These energies differ by one order in magnitude, 
which implies that a polaron dominates the charge transport. 
A sign of the thermopower of the polycrystals changes 
from negative to positive at 500~K, indicating that holes are excited thermally 
to decrease the magnitude of thermopower.
\end{abstract}

\date{\today}

\maketitle

\section{Introduction}
Thermoelectric materials, which convert heat into electricity and 
vice versa through the thermoelectric phenomena in solids, 
have recently attracted a renewed interest 
as a promising energy-conversion technology 
that is friendly to the environment. 
The conversion efficiency of a thermoelectric material 
is characterized by the figure of merit 
$Z = S^2/\rho \kappa$, where $S$, $\rho$ and $\kappa$ 
are the thermopower, the resistivity, 
and the thermal conductivity, respectively. 

Recently Terasaki {\it et al}.\cite{terasaki} found 
that a layered cobalt oxide NaCo$_2$O$_4$ shows a large thermopower 
of 100 $\mu$V/K and a low resistivity of 200 $\mu\Omega$cm 
at room temperature, 
and they proposed that the strong correlation plays an important 
role in the enhancement of the thermoelectric properties. 
Based on this proposal, other layered cobalt oxides 
Bi-Sr-Co-O\cite{itoh,funahashi} and Ca-Co-O\cite{co349,miyazaki,co225} 
have been studied, and they are found to be 
potential thermoelectric materials as well as NaCo$_2$O$_4$. 
In particular, Funahashi {\it et al}.\cite{co225} reported 
the maximum of $ZT$ is larger than 1.2 for Ca-Co-O, 
which means that the layered cobalt oxides might be the best group 
among all the thermoelectric materials.

These layered cobalt oxides are, however, all p-type materials, 
and an n-type cobalt oxide has not been found yet. 
Thermoelectric devices are made from a pair of p and n-type 
thermoelectric materials, and we tried to find an n-type cobalt oxide. 
The Co ion in NaCo$_2$O$_4$ is at the center of 
an oxygen octahedron, and its average valence is +3.5
with the electron configuration of $(3d)^{5.5}$,
which corresponds to 0.5 hole in the $t_{2g}$ bands per Co site.  
In NaCo$_2$O$_4$, the octahedron is edge-shared, 
where the $t_{2g}$ bands of Co 3$d$ directly overlap each other. 
By contrast, electrons doped in an insulating cobalt oxide 
made of Co$^{3+}$ will be in the $e_g$ bands.
Since the $e_g$ bands are strongly hybridized with O $2p$,
the conduction-band width is largest when the bonding angle of Co-O-Co
is 180$^{\circ}$.
Thus a cobalt oxide in a corner-shared oxygen octahedron
will be a good n-type conductor.

\begin{figure}
 \includegraphics[width=7cm,clip]{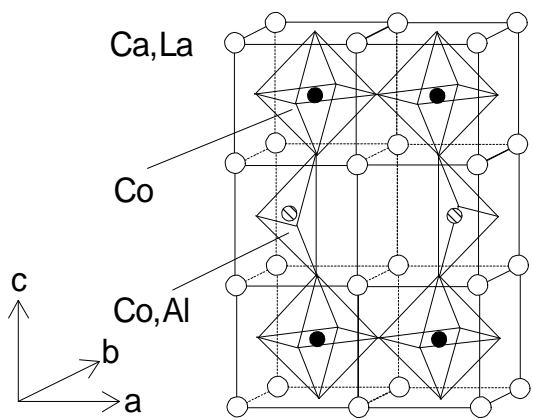}
 \caption{
 Schematic view of the brownmillerite structure.
 }
\end{figure}

Accordingly we searched for a cobalt oxide in a corner-shared 
oxygen octahedron, 
and eventually found Ca$_2$(Co,Al)$_2$O$_5$\cite{kouzou}, 
which has a brownmillerite structure with Co$^{3+}$. 
The brownmillerite structure is schematically shown in Fig. 1. 
Big white circles represent Ca and La, black circles Co, 
shaded circles Al and Co. 
This structure is composed of an alternate stack of the CoO$_2$ square 
lattice and the (Co,Al)O chain along the $c$ direction, 
and can be regarded as an oxide-deficient perovskite structure. 
In this paper we report on preparation of polycrystalline 
and single-crystal samples, and on measurements and analyses of 
their thermoelectric properties.

\section{Experimental}
Polycrystalline samples of Ca$_{2-y}$La$_y$CoAlO$_5$ 
($y=$0, 0.05, 0.1 and 0.15) and 
Ca$_{1.95}$La$_{0.05}$Co$_{2-x}$Al$_x$O$_5$ 
($x=$0.4, 0.5, 0.6, and 0.7)~were prepared by a solid-state reaction. 
Stoichiometric amounts of CaCO$_3$, La$_2$O$_3$, Co$_3$O$_4$ 
and Al$_2$O$_3$ were mixed, and the mixture was calcined 
at 900$^{\circ}$C for 12h in air. 
The product was finely ground, pressed into a pellet, 
and sintered at 960$^{\circ}$C for 12h in air.

Single-crystal samples were prepared by a Bi$_2$O$_3$/PbO flux method. 
Bi$_2$O$_3$, PbO, CaCO$_3$ and Co$_3$O$_4$ powders were 
mixed with a cationic composition of Bi:Pb:Ca:Co=1.3:0.7:3:2, 
and was heated at 1100$^{\circ}$C for 5h, and then slowly 
cooled down to 700$^{\circ}$C at a rate of 8$^{\circ}$C/h in air 
using an Al$_2$O$_3$ crucible (Al was supplied from the crucible). 
The composition ratio of a single-crystal sample measured by 
energy dispersive X-ray analysis (EDX) is Ca:Co:Al$=$2:1.3:0.7. 

The X-ray diffraction (XRD) of the sample was measured using 
a standard diffractometer with Fe K$_\alpha$ radiation as an X-ray source 
in the $\theta-2\theta$ scan mode. 
The resistivity was measured using a four-probe method, 
below room temperature in a liquid He cryostat, 
and above room temperature in an electric furnace. 
The thermopower was measured using a steady-state technique, 
below room temperature in a liquid He cryostat, 
and above room temperature in an electric furnace. 
A temperature gradient of 1~K/cm was generated 
by a small resistive heater pasted on one edge of a sample, 
and was monitored by a differential thermocouple made 
of copper-constantan below room temperature, 
and by that of platinum-rhodium above room temperature. 
The thermopower of the voltage leads was carefully subtracted. 
The thermal conductivity was measured from 30 to 280~K 
using a steady-state technique in a closed refrigerator 
pumped down to 10$^{-6}$~Torr. 
The temperature gradient was monitored using a differential 
thermocouple made of chromel-constantan.

\begin{figure}
 \includegraphics[width=8cm,clip]{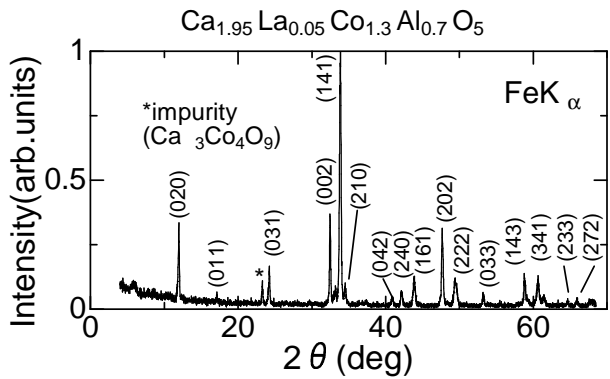}
 \caption{
 X-ray diffraction pattern of Ca$_{1.95}$La$_{0.05}$Co$_{1.3}$Al$_{0.7}$O$_5$. 
 }
\end{figure}

\section{Results and Discussion}
Figure 2 shows an XRD pattern of a polycrystalline sample
Ca$_{1.95}$La$_{0.05}$Co$_{1.3}$Al$_{0.7}$O$_5$ 
A small amount (at most 7\%)of impurity phase 
(mainly Ca$_3$Co$_4$O$_9$)\cite{co349} 
is seen, but almost all the peaks are indexed as the brownmillerite phase. 
The lattice parameters were obtained to be $a=$5.28~\AA,
$b=$5.52~\AA, and $c=$14.66~\AA, which is consistent 
with a 4-cycle X-ray diffraction of a single crystal. 
We will use the composition as the nominal composition in the present paper, 
though the real composition is different from the nominal composition.
However, the difference is small within less than 10\% accuracy:
for example a nominal La [Al] content of 0.05 [0.5]
corresponds to 0.053 [0.54] for the real composition at most.
We further note that the impurity phase of Ca$_3$Co$_4$O$_9$
shows positive and nearly temperature-independent thermopower
above 100 K \cite{co349}, which would not seriously affect 
the temperature and the composition dependences of $\rho$ and $S$,
or the discussion in the present paper.

\begin{figure}
 \includegraphics[width=7cm,clip]{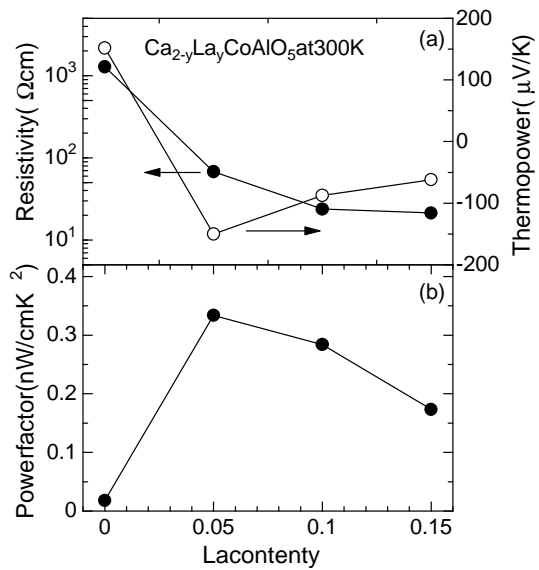}
 \caption{
 (a) Resistivity and the thermopower of Ca$_{2-y}$La$_y$CoAlO$_5$ at 300 K
 and (b) power factor of Ca$_{2-y}$La$_y$CoAlO$_5$ at 300K.
 }
\end{figure}

Figure 3(a) shows $\rho$ and $S$ of polycrystalline 
Ca$_{2-y}$La$_y$CoAlO$_5$ at room temperature. 
With increasing La content $y$, $\rho$ decreases systematically, 
and the sign of $S$ changes from positive to negative. 
This indicates that the substitution of La$^{3+}$ for Ca$^{2+}$ 
supplies an electron. 
Figure 3(b) shows the power factor $S^{2}/\rho$ 
corresponding to the data in Fig 2(a). 
The power factor takes a maximum at $y$=0.05, and accordingly 
we fix $y$ to be 0.05 to see the Al-content dependence.

Figure 4(a) shows $\rho$ of polycrystalline samples 
of Ca$_{1.95}$La$_{0.05}$Co$_{2-x}$Al$_x$O$_5$, 
together with the in-plane (parallel to the $ab$ direction) 
and the out-of-plane (parallel to the $c$ direction) 
resistivity of a single-crystal sample of Ca$_2$Co$_{1.3}$Al$_{0.7}$O$_y$. 
At 750~K, the magnitude of resistivity 
decreases down to 10-20 m$\Omega$cm for polycrystalline samples
and the in-plane direction of the single crystals. 
This magnitude of resistivity is the lowest, to our knowledge, 
in brownmillerite oxides \cite{brown1,brown2}. 
The temperature dependence is described by an activation-type conduction, 
where the activation energy for resistivity (E$_\rho$) is approximately 
0.2 eV for all the samples.
As for anisotropy, the out-of-plane resistivity is 100 times
higher than that for the in-plane resistivity,
implying that the layered structure is realized,
as is expected from the brownmillerite phase. 
This means that the conduction path of a single crystal sample 
is two-dimensional, and the conduction path of the polycrystalline
samples is mainly determined by the in-plane conduction paths. 
Then the  resistivity of the polycrystalline samples is
higher than the in-plane resistivity of the single crystal sample,
while the temperature dependences are similar,
as is also seen in the resistivity of high-temperature superconductors.
The resistivity increases with increasing Al content $x$, 
which indicates that a Co ion is responsible for the electric conduction
and an Al ion acts as an impurity.
We should note that the brownmillerite structure was not synthesized 
for $x\le 0.4$, which means that an Al ion is essential to stabilizing
the brownmillerite structure at ambient pressure.

\begin{figure}
 \includegraphics[width=8cm,clip]{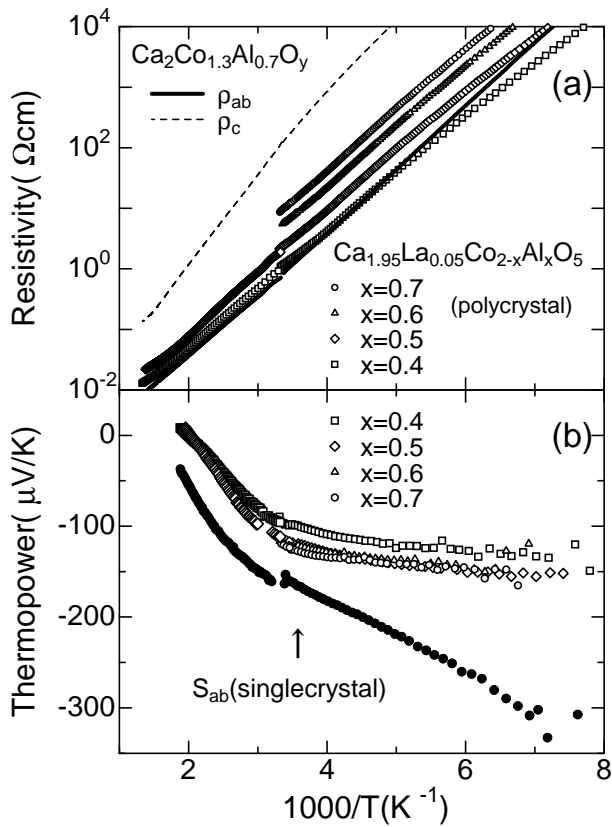}
 \caption{
 (a) Resistivity and (b) thermopower 
 of polycrystalline Ca$_{1.95}$La$_{0.05}$Co$_{2-x}$Al$_x$O$_5$
 and single-crystal Ca$_2$Co$_{1.3}$Al$_{0.7}$O$_y$ 
}
\end{figure}

Figure 4(b) shows the thermopower of the prepared samples. 
The in-plane thermopower of the single-crystal sample 
is $-$150~$\mu$V/K, 
and that of polycrystalline samples is about $-$100~$\mu$V/K at 300 K. 
This means that the carrier concentration of the single-crystal sample
is smaller than that of the polycrystalline samples, 
because no donor ion such as La is included in the single crystal.
The source of the carrier is not clear at present,
and the oxygen deficiency or a small inclusion of Bi$^{3+}$
are possible candidates. 

It should be emphasized that the magnitude and the temperature 
dependence of the thermopower of polycrystalline samples are nearly independent of $x$. 
Considering the $x$ dependence of $\rho$ and $S$, 
we can conclude that an Al ion acts as a scattering center 
(disorder) for electric conduction,
although it stabilizes the brownmillerite structure
at ambient pressure \cite{kouzou}.
Consequently carriers are likely to conduct on CoO$_2$ square 
lattice, not on the (Co,Al)O chain. 
Below 350 K, the temperature dependence of the thermopower is described 
by an activation-type conduction with the activation energy for 
thermopower ($E_s$) of 0.01 eV for the polycrystalline samples, 
and 0.04 eV for the single-crystal sample. 
$E_\rho (=0.2~{\rm eV}) \gg E_s (=0.01-0.04~{\rm eV})$ 
implies a polaron conduction, 
where $E_\rho$ and $E_s$ are described as $E_\rho=E_F-E_c+W_H$ 
and $E_s=E_F-E_c$ ($E_c$: the energy of bottom of conduction band, 
$E_F$: Fermi energy, $W_H$: hopping energy that contributes 
to the change of mobility by temperature)\cite{mott}. 
Such a material is useful for the high-temperature 
thermoelectric application \cite{wood}, 
because a rapid decrease in $\rho$ and a gradual decrease in $S$ 
will occur with increasing temperature.

For the present material, however, the high temperature thermopower is quite small. 
Above 350~K it deviates from the activation type, 
and the sign of the thermopower of the polycrystalline samples changes 
from negative to positive at 500 K. 
A similar tendency was observed for the single-crystal sample, 
though a sign change was not yet seen below 550~K. 
As a result the thermoelectric performance rapidly 
decreases above 450~K.
Very recently Ueno {\it et al.} have found a similar sign change 
of the thermopower in the brownmillerite Ca-Co-Al-O 
thin film \cite{ueno}.

Figure 5 shows the thermal conductivity of polycrystalline samples. 
The magnitude is about 20~mW/cmK, which is relatively low, 
possibly due to a solid solution of Co and Al. 
Using $\kappa$=20 mW/cmK, the figure of merit for $x=0.4$
is estimated to be of the order of 10$^{-6}$ K$^{-1}$ at 300 K,
which is three orders of magnitude smaller than that of the conventional 
thermoelectric materials.
By examining $\rho$, $S$ and $\kappa$ individually, we find $\rho$ is
hopelessly high, while $S$ and $\kappa$ are comparable.
This is due to the polaron conduction where the mobility is
exponentially lowered with decreasing temperature.

\begin{figure}
 \includegraphics[width=8cm,clip]{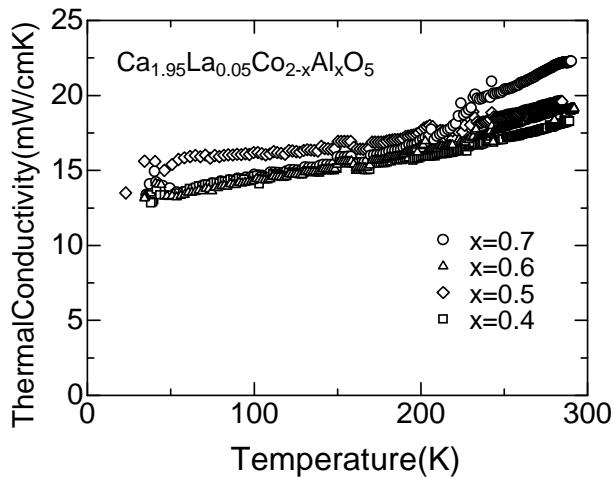}
\caption{
 Thermal conductivity of polycrystalline 
 Ca$_{1.95}$La$_{0.05}$Co$_{2-x}$Al$_x$O$_5$  
}
\end{figure}

Let us discuss the sign change of the thermopower at 500 K. 
We should note that the electronic states (and the crystal structure)
of Ca$_2$(Co,Al)$_2$O$_5$ resemble 
those of the cubic perovskite LaCoO$_3$,
which is a p-type conductor at all temperatures\cite{laco}.
Thus it is very likely that holes are thermally excited
also in Ca$_2$(Co,Al)$_2$O$_5$ at high temperature, 
which will compensate the doped electrons 
to decrease the magnitude of the thermopower.
In general, the square lattice as seen in the brownmillerite
is often symmetric to electron and hole doping 
(which is exact in the tight-binding approximation) \cite{mao, mackenzie}.
This means that electrons and holes are equally excited 
at high temperature to give zero thermopower.
Thus the electron-hole symmetry is unsuitable for
good thermoelectrics, which might be the reason 
most of the perovskite related oxides are not 
good thermoelectric materials.

\section{Summary}
We successfully prepared an n-type Co oxide material for the first time, 
although the thermoelectric properties are not satisfactory.
The figure of merit at 300 K is of the order of 
$10^{-6}$ K$^{-1}$, 
the low value of which comes from the high resistivity. 
The high resistivity is ascribed to the 
polaron conduction suggested by the fact
that $E_\rho$ is one order of magnitude larger than $E_s$.
The sign change of the thermopower represents 
thermal excitation of holes at high temperatures. 
Thus it would be less advantageous to design an n-type thermoelectric
material by using Co oxides with a corner-shared oxygen 
octahedron network.

\section{Acknowledgements}
The authors appreciate K. Takahata for fruitful discussions and valuable comments. 
We would also like to thank R. Kitawaki and T. Itoh for collaboration.


\begin{thebibliography}{9}
\bibitem{terasaki}
	I. Terasaki, Y. Sasago and K. Uchinokura: 
	Phys. Rev. {\bf B56} (1997) R12685.
 \bibitem{itoh}
	T. Itoh, and I. Terasaki: 
	Jpn. J. Appl. Phys. {\bf 39} (2000) 6658.
 \bibitem{funahashi} 
	R. Funahashi and I. Matsubara: 
	Appl. Phys. Lett. {\bf 79} (2001) 362.
 \bibitem{miyazaki}
	Y. Miyazaki, K. Kudo, M. Akoshima, Y. Ono, Y. Koike and
	T. Kajitani:
	Jpn. J. Appl. Phys. {\bf 39} (2000) L531.
 \bibitem{co349}
	A. C. Masset, C. Michel, A. Maignan, M. Hervieu, 
	O. Toulemonde, F. Studer, B. Raveau, and J. Hejtmanek: 
	Phys. Rev. {\bf B62} (2000) 166.
 \bibitem{co225}
	R. Funahashi, I. Matsubara, H. Ikuta, T. Takeuchi, 
	U. Mizutani, and S. Sodeoka: 
	Jpn. J. Appl. Phys. {\bf 39} (2000) L1127.
 \bibitem{kouzou}
	J. Y. Lee, J. S. Swinnea, and H. Steinfink: 
	Acta Cryst. {\bf C47} (1991) 1532.
 \bibitem{brown1}
	G. B. Zhang, and D. M. Smyth: 
	Solid State Ionics {\bf 82} (1995) 161.
 \bibitem{brown2}
	M. Schwartz, B. F. Link and A. F. Sammells: 
	J. Electrochem. Soc. {\bf 140} (1993) L62.
 \bibitem{mott}
	N. F. Mott: 
	Metal-Insulator Transitions Second Edition 
	(1990, Taylor and Francis).
 \bibitem{wood}
	C. Wood and D. Emin:
	Phys. Rev. {\bf B29} (1984) 4582
 \bibitem{ueno}
	M. Ueno, T. Terada, T. Yoshida, N. Matsunami,
	and Y. Takai:
	Annual Meeting of Japan Society of Applied Physics, 
	September 2001 (in Japanese)
 \bibitem{laco}
	M. A. Se\~nar\'{\i}s-Rodr\'{\i}guez, and J. B. Goodenough: 
	J. Solid. State. Chem. {\bf 116} (1995) 224.
 \bibitem{mao}
	W. Mao and K. S. Bedell:
	Phys. Rev. {\bf B59} (1999) R15590.
 \bibitem{mackenzie}
	J. Merino and R. H. McKenzie:
	Phys. Rev. {\bf B61} (2000) 7996.
\end{thebibliography}
\end{document}